\documentclass{ws-procs9x6}

\begin{document}

\title{Baryon Structure\footnote{\uppercase{T}his work is partly supported by 
\uppercase{COFINANZIAMENTO MURST-PRIN-01}}}

\author{M.~RADICI}

\address{Istituto Nazionale di Fisica Nucleare, Sezione di Pavia, and \\
Dipartimento di Fisica Nucleare e Teorica, Universit\`a di Pavia, \\
via Bassi 6, \\ 
27100 Pavia, Italy\\ 
E-mail: radici@pv.infn.it}


\maketitle

\abstracts{A review of the theoretical activity in Italy in the research field of
Hadronic Physics is given. Specific focus is put on phenomenological models based on
the effective degrees of freedom of constituent quarks, on parton distributions in 
hard processes in the Bjorken limit and on the possibility of linking the two concepts 
via evolution equations. A brief introduction is given also about the socalled 
generalized parton distributions. }

%%%%%%%%%%%%%%%%%%%%%%%  Introduction %%%%%%%%%%%%%%%%%%%%%%%%%%%

\section{Introduction}
\label{sec:intro}

The Quantum Chromodynamics (QCD) is the commonly accepted theory of strong interactions
of elementary particles. It displays two fundamental properties: the asymptotic freedom
at very short distances and the confinement of quarks and gluons inside the 
observed colorless hadrons. While the former allows for spectacular precision tests of
the theory in the framework of the socalled perturbative QCD (pQCD), the latter yet
cannot be explained from first principles. The reason is that the confinement happens at
the hadronic scale (typically, the nucleon size) where the running coupling, $\alpha_s$,
is so large to forbid any perturbative approach. There are several ways to attack the 
problem: formulating QCD sum rules, solving the theory on the lattice, etc.. Here, I 
will focus on the attempts to explore the nonperturbative regime using ans\"atze induced 
from the fundamental symmetries of QCD itself. More explicitly, this kind of activity, 
conventionally included in the socalled Hadronic Physics, can be reformulated as the 
study of confined systems of strongly interacting quarks and gluons. From this point of 
view, the overlap with the know-how of the Nuclear Physics community is very large, 
since the hadron can be considered as a strongly correlated many-body system. 

In the following, I will review the theoretical contributions in Italy to this research
field. In Sec.~\ref{sec:cqm} I will focus on different formulations of low-energy 
phenomenological models based on the effective degrees of freedom of constituent quarks,
with emphasys on the covariance and gauge invariance of such models. In
Sec.~\ref{sec:dis} I will concentrate on the analysis of hard processes at a higher
scale, in a typical regime of Deep-Inelastic Scattering (DIS), with the aim of 
exploring the (spin) properties of partons inside the hadrons and, possibly, of
establishing a link between the concept of constituent quark (around $\Lambda_{\rm
QCD}$) and current quark (in the DIS regime) through the evolution equations. Finally, 
in Sec.~\ref{sec:gpd}, I will consider the socalled Generalized Parton Distributions 
(GPD), a very promising formalism which unifies both exclusive and (semi)inclusive 
hard processes, as it may offer a gauge invariant solution to the puzzle of the nucleon 
spin decomposition into the spins of its elementary constituents. 

%%%%%%%%%%%%%%%%%%%%%%  Constituent Quark Models  %%%%%%%%%%%%%%%%%%%%%%%%

\section{Constituent Quark Models}
\label{sec:cqm}

Constituent Quark Models (CQM) are described by hamiltonians involving effective degrees 
of freedom, the constituent quarks, which are massive quasi-particles moving in a 
background field generated by gluons. Confinement is realized by a long-range 
phenomenological potential $V_{\rm conf}$, while the short-range interaction 
$V_{\rm qq}$ is treated perturbatively. The spectrum is obtained by diagonalizing the 
hamiltonian. Since this is not sufficient to discriminate among different dynamical 
models, hadron electromagnetic observables are also deduced by computing matrix elements 
of the electromagnetic current operator between the obtained eigenfunctions. 

A typical example is represented by the socalled hypercentral CQM of baryons, where the 
unperturbed hamiltonian is given by a linear and a coulombic contributions in terms of 
the hyperradius $x=\sqrt{\rho^2+\lambda^2}$, with $\rho, \lambda$, the usual Jacobi 
coordinates of the three-quark system. Such a three-body hypercentral potential 
satisfactorily reproduces the SU(6) gross structure of the light-baryon spectrum using 
just two free parameters. The further hyperfine splittings were interpreted, in a first 
version of the model, by adopting for $V_{\rm qq}$ the spin-spin part of the hyperfine 
interaction induced by a one-gluon-exchange (OGE) mechanism~\cite{ge1} (employing two
new parameters). However, despite the introduction of relativistic corrections both of 
kinematical and dynamical type to the current operator~\cite{ge-talk}, several 
drawbacks arise. In particular, the wrong ordering in the spectrum of the Roper 
resonance $J^P=\textstyle{1\over 2}^-$ with respect to its positive-parity partner, and
the impossibility of finding a consistent parametrization for the root-mean-squared 
(rms) charge radius $\langle r^2 \rangle_{\rm ch}$ and the nucleon (N) electromagnetic 
form factors at the same time. A new version of the hypercentral CQM has been recently 
released~\cite{ge2}, which implements $V_{\rm OGE}$ by adding also an explicit isospin 
dependence (and introducing four new parameters), as suggested also by other models of 
$V_{\rm qq}$ (see Sec.\ref{subsec:pf} in the following). This SU(6)-breaking feature is 
crucial to get a much more accurate description of the spectrum. 

However, the hypercentral CQM is based on a nonrelativistic description, which is known
to be inadequate for constituent quarks that are essentially relativistic particles. The 
question arises why these kinds of nonrelativistic CQM reproduce several hadron 
properties at the quantitative level. A possible answer stems from the socalled General 
Parametrization (GP) method, according to which the (QCD) exact matrix element of an 
hermitean operator can be represented by a linear parametrization of the complete set 
of spin-flavor operators of three constituent quarks. 
%acting on the Hilbert space of three valence quarks. 
The key observation is that there is a hierarchy in the size of the parameters, the 
nonrelativistic CQM retaining just the first largest ones. Several applications of the 
GP method lead to interesting results~\cite{ge3,ge4,ge5,ge6}. The GP method considers 
matrix elements in the hadron rest frame, thus loosing the property of covariance but 
keeping Lorentz invariance. 

%Moreover, the masses and average momenta of constituent quarks 
%are of the same order of magnitude as $\Lambda_{\rm QCD}$. A consistent treatment 
%requires, therefore, a relativistic covariant quantum-mechanical formulation of the
%dynamics of constituent quarks. 
%In general, in CQM the constituent quarks can have momenta much larger than their masses. 
%Therefore, a consistent description of their dynamics requires a relativistic covariant 
%quantum-mechanical formulation. 
Another series of works consists in covariant approaches based on a given hamiltonian 
with a finite number of degrees of freedom. Along this line of investigation, a 
relativistic quantum-mechanical formulation is only possible in terms of unitary 
representations of the Poincar\'e group. There are at least three unitarily equivalent 
ways to do it: the instant form (IF), the light-front form (LF) and the point form (PF) 
of dynamics. They can be distinguished by looking at the different ways of implementing 
the interaction into a free covariant theory while keeping the covariance and the Lie 
algebra of the 10 Poincar\'e generators, the socalled Bakamjian-Thomas (BT) construction. 
The instant form is the usual context in which all relativistic field theories, 
including QCD, are developed. In the following, I will review results obtained by the LF 
and PF descriptions of relativistic CQM, respectively.

%%%%%%%%%%%%%%%%%%%%%%%%%%%%% Light-Front %%%%%%%%%%%%%%%%%%%%%%%%%%%%%%%%%%

\subsection{Front Form Description}
\label{subsec:lf}

The starting point is the eigenvalue problem for a Poincar\'e- (${\mathcal P}$-) 
covariant mass operator $M_0+V$, with $M_0$ the free mass operator. If the interaction 
$V$ is independent of the total momentum ${\bf P}_{\rm tot}$ of the system and commutes 
with the LF spin operator, then the eigenvalue 
problem can be rewritten in terms of an intrinsic relativistic kinetic operator 
(involving the $N$ constituent momenta ${\bf k}_i$ such that $\sum_i^N {\bf k}_i = 0$) 
and of the potential ${\mathcal V} = {\mathcal R} V {\mathcal R}^\dagger$, where
${\mathcal R}$ is the Melosh rotation connecting the spins in the (IF) standard 
representation to the LF representation. When calculating electromagnetic observables, 
the current operator $J^\mu$ must be a conserved current and must fulfill the socalled 
extended ${\mathcal P}$ covariance, i.e. it must be ${\mathcal P}$-covariant and it must 
respect parity $(P)$ and time-reversal $(T)$ invariance. Then, the form factors can be 
expressed in terms of matrix elements of only $J^+$ and the socalled angular condition 
is satisfied, i.e. a relation between the current matrix elements must hold such that 
the number of form factors corresponds to the number of $J^+$  independent matrix 
elements. If, moreover, the Drell-Yan condition for the momentum transfer is satisfied, 
i.e. $q^+=0$, the momentum conservation on the light-cone implies, for systems with 
total spin $J\leq {\textstyle{1\over 2}}$, a suppression of many-body contributions to 
the electromagnetic vertex, corresponding, in field theory, to Feynman diagrams 
involving nonvalence $q\bar q$ pairs. Thus, it is legitimate to use the Impulse 
Approximation (IA) for the current operator $J^\mu_{\rm IA}$.
%\begin{equation}
%J^\mu_{\rm IA} = \sum_i^N \; \left[ e_i \, f_1^i(Q^2) \, \gamma^\mu + i \, \kappa_i \, 
%f_2^i(Q^2) \, \frac{\sigma^{\mu\nu} q_\nu}{2m_i} \right] \; ,
%\label{eq:iacurr}
%\end{equation}
%where $e_i, m_i, \kappa_i,$ are the charge, mass and anomalous magnetic moment of each
%of the $N$ constituent quarks and $f_{1(2)}^i$ are their Dirac (Pauli) form factors as
%functions of the invariant momentum transfer $Q^2=-q^2$. Considering just the $u,d$ 
%flavors, $f_{1(2)}$ for the proton and the neutron are described by 12 parameters that 
%can be constrained by fitting the N and $\pi$ elastic form factors, while transitions 
%to various N$^*$ resonances can be predicted in a parameter-free way.
Introducing parametrized form factors for the constituent quarks, a global fit to the
spectrum of light baryons and to the N and $\pi$ elastic form factors is achieved, which
constrains the parameters in order to make predictions for the transitions to the various
N$^*$ resonances. Using for ${\mathcal V}$ the Capstick-Isgur potential based on a 
linear confining term and on an SU(6)-breaking effective gluon-exchange contribution, 
which includes a Coulomb-like part, a spin-spin part responsible for the hyperfine 
splittings, and a tensor part, a successful and consistent picture of light hadrons is
obtained~\cite{rmII1}. 

However, the use of $J^\mu_{\rm IA}$ breaks the angular condition. This is particular
evident in the case of the N$\rightarrow \Delta$ transition, where three form factors
(electric, $G_E$; magnetic, $G_M$; coulombic, $G_C$) correspond to four independent
matrix elements. Usually, three out of the latter ones are chosen to be independent and 
the fourth one is forced to fulfill the angular condition, but this selection is 
arbitrary reflecting in very different results for the various multipoles contributing 
to the transition~\cite{rmII2}. In the case of a system with total spin 
$J\leq {\textstyle{1\over 2}}$, there is a perfect correspondence between form factors 
and $J^+_{\rm IA}$ matrix elements. However, the angular condition is still broken, or,
equivalently, the rotational covariance of the current operator is lost. For the example 
of the N elastic form factors, the $G_M^{\rm N}(Q^2)$ can be deduced from different 
components of $J^\mu_{\rm IA}$ obtaining two different results~\cite{rmIII1}. Moreover,
it seems not possible to reproduce at the same time the two main evidences for the
breaking of SU(6) symmetry: the nonvanishing neutron electric form factor 
$G_E^{\rm n}(Q^2)$ and the deviation of the ratio $G_M^{\rm p}(Q^2)/G_M^{\rm n}(Q^2)$ 
from the value -1.5. Cardarelli and Simula~\cite{rmIII1} suggest a common recipe to these
problems. The central idea is that for a system with $J\leq {\textstyle{1\over 2}}$ in 
the $q^+=0$ frame the induced loss of rotational covariance implies that the 
$J^\mu_{\rm IA}$ matrix elements depend upon the choice of a null 4-vector $\omega$ 
perpendicular to the LF null plane $x^+=0$. The constituent quark momenta can then be 
reparametrized as to be all on shell, as demanded by the IA. Consequently, the 
covariant decomposition of the current operator contains new spurious structures 
explicitly depending on $\omega$. Overcoming the angular
condition problem means selecting appropriate components of $J^\mu_{\rm IA}$ such that 
the calculated observables do not depend upon spurious contributions, as they have to. In
this way, good results are obtained for the electromagnetic observables of pion and
nucleon, while the situation is less clear in the case of N inelastic transitions 
because of the scarce amount of data available~\cite{rmIII2}. 

For spin-1 systems it is known that also for $q^+=0$ the contribution of diagrams 
involving $q\bar q$ pairs does not vanish, so that the question arises if an IA in a LF
model with fixed number of degrees of freedom is possible. Melikhov and 
Simula~\cite{rmIII3} show that this is possible by extending the previous analysis to the
tensor structure of the deuteron electromagnetic vertex, with the interesting outcome
that also ``good'' components for the initial and final LF polarization vectors of the 
spin-1 system must be selected in order to get matrix elements free from spurious
contributions. 

An alternative solution to the problem of rotational covariance is discussed by Lev,
Pace and Salm\`e~\cite{rmII2}. From the fact that LF Lorentz boosts are interaction 
free the wave function is usually split in a cm and an intrinsic part. Similarly, 
taking matrix elements of the exact current operator $J^\mu$ on projected N states in 
the Breit frame with ${\bf q}\parallel \hat z$, it is possible to construct a current 
operator, acting only on the intrinsic Hilbert subspace, that is extended 
${\mathcal P}$-covariant, hermitean, conserved, and fulfills charge normalization. For 
all the properties to hold, the choice of the frame is crucial, because the rotational 
covariance around $\hat z$ is sufficient to grant all the other features. Since in the 
LF these rotations are interaction free, the current operator can be built starting from 
$J^\mu_{\rm IA}$, while the angular condition is still satisfied. For the example of the
deuteron, improvement is obtained for both the magnetic and quadrupole momenta,
indicating the importance of relativistic covariance even at $Q^2=0$; the three 
electromagnetic form factors can be unambiguously calculated, even if the results are
still affected by model dependence in the choice of the N-N interaction and of the N form
factors~\cite{rmII2}.

As already stressed, the above recipe of building an ``intrinsic'' covariant operator 
relies on the choice of the Breit frame with ${\bf q}\parallel \hat z$, or equivalently 
for ${\bf q}_\perp = 0$. Therefore, the pair diagrams do contribute to one-loop 
electromagnetic interactions in a covariant field-theoretical approach. In 
Ref.~\cite{rmIII4}, Simula discusses the issue of rotational covariance by comparing it 
with the previous recipe based on the elimination of spurious contributions. By 
analyzing the form factors of the pion and of the deuteron, he shows that 
the two approaches are inequivalent, since the LF Lorentz transformation from the 
$q^+=0$ frame to the ${\bf q}_\perp =0$ one is interaction dependent. The claim is that 
the pair diagrams can be safely neglected in the $q^+=0$ frame, but they are important at
$q^+\neq 0$ for light hadrons, disappearing in the heavy-quark limit. In another
paper~\cite{rmII3}, de Melo, Frederico, Pace and Salm\`e analyze the electromagnetic 
form factor of a pion-like $q\bar q$ bound state by building a covariant 
model of the Bethe-Salpeter amplitude with a symmetric vertex function in momentum 
space, deduced from a lagrangian with pseudoscalar pion-quark coupling at the one-loop 
level. By parametrizing the Breit frame with an angle $\alpha$ indicating the position 
of the momentum tranfer ${\bf q}$ in the $\widehat{xz}$ plane, they study the relative
weight of the spectator and pair diagrams with varying $\alpha$ between the boundary
positions $\alpha =0$ ($q^+=0$) and $\alpha =90^{\rm o}$ ($q^+=q$), recovering
essentially the same results as in Ref.~\cite{rmIII4}. However, by rescaling the 
constituent quark masses to the N masses such that the pion-like system becomes a 
weakly relativistic spin-0 system with approximately the deuteron mass, the interesting 
result emerges that the pair diagram can be safely neglected for $Q^2\leq 1$ and for any 
$\alpha$, i.e. in any Breit frame with $q_y=0$.

%%%%%%%%%%%%%%%%%%%%%%%%%%%%% Point-Form %%%%%%%%%%%%%%%%%%%%%%%%%%%%%%%%%%

\subsection{Point Form Description}
\label{subsec:pf}

In the PF realization of relativistic quantum mechanics only the four-momentum operator
$P^\mu$, generating the space-time translations, contains the interaction. The theory is 
thus manifestly covariant and the current matrix elements can be accurately calculated 
in any frame. It is convenient to introduce the socalled velocity states, defined in such
a way that, under a Lorentz transformation $\Lambda$, the momenta and the spins of the
constituents are rotated by the same Wigner rotation $R_W$; contrary to the LF
formalism, it is then possible to couple orbital angular momenta and spins in the same 
way as it is done in a nonrelativistic theory. By construction, the velocity states are 
eigenfunctions of $M_0$ and of the free momentum operator $P^\mu_0 = M_0 V^\mu$. 
Interactions are introduced in the momentum operator as 
$P^\mu = P_0^\mu + P_I^\mu \equiv M V^\mu$, where the mass operator is $M = M_0 + V$. 
The BT construction is realized if $M$ is a scalar (or, equivalently, $V$ is 
rotationally invariant) and commutes with $V^\mu$ (i.e., $V$ is a local interaction); 
the Lie algebra of the Poincar\'e group, then, takes the form of the eigenvalue problem 
$M |\psi\rangle = m|\psi\rangle$ for a system of mass $m$ made up by $N$ constituents. 

If the current operator $J^\mu$ satisfies the same Lie algebra, it can be shown to be an 
irreducible tensor operator of the Poincar\'e group. Then, a generalized Wigner-Eckhart 
theorem implies that if the matrix elements are taken on the eigenstates of $P^\mu$, 
they can be written as the product of covariant objects (the Clebsh-Gordan coefficients 
of the Poincar\'e group) and invariants $G^\mu(Q^2)$ (the "PF" form factors). In the 
Breit frame with ${\bf q}\parallel \hat z$, the Clebsh-Gordan coefficients become the 
identity, so that the $G^\mu(Q^2)$ can be deduced by directly computing the matrix 
elements of $J^\mu$ in this frame. Recalling the discussion developed in 
Sec.~\ref{subsec:lf}, the angular condition is here automatically satisfied if $J^\mu$ 
is also $P-$ and $T-$reversal invariant: for a system with total angular momentum $J$ 
there are only $2J+1$ invariant "PF" form factors. 

By projecting the initial and final states in Breit frame upon the corresponding 
velocity states with $v_i=-Q/2m$ and $v_f=Q/2m$, it is possible 
to directly calculate the needed matrix elements of $J^\mu$ using the elementary 
degrees of freedom. As in the LF case, the IA is adopted but the
constituents are here considered point-like. The presence of velocity states implies 
that the momentum conservation constraint cannot be written in close form. In 
other words, the impulse $q=mv_f - mv_i$ delivered to the system is different from the 
one absorbed by the active constituent. For this reason, this IA in PF is called the 
Point-Form Spectator Approximation (PFSA)~\cite{GBE1}. The role played by the velocity
dependent boosts is crucial in determining the correct $Q^2$ fall-off of the form 
factors. 

In Refs.~\cite{GBE1,GBE2,GBE3} results are shown for the PFSA in a relativistic CQM 
based on the socalled Goldstone-Boson Exchange (GBE) quark-quark interaction 
$V_{\rm GBE}$, which is dictated by the spontaneous breaking of QCD chiral symmetry at 
low energy. As such, the model is applicable below the energy threshold of chiral 
symmetry restoration, in any case for $Q^2\lesssim 3 \div 4$ (GeV/$c)^2$. At variance 
with $V_{\rm OGE}$, $V_{\rm GBE}$ explicitly depends on flavor; its peculiar structure 
is responsible for the very good reproduction of the low-lying light and strange baryon
spectra using just five parameters. In particular, it offers a natural solution to the 
problem of the correct position of the Roper resonance energy level. Covariant 
calculations of electroweak N form factors have been performed in the PFSA obtaining an 
overall very consistent picture~\cite{GBE3}. The agreement is particularly impressive if 
it is considered that 
point-like constituent quarks are used and that only the GBE model wave functions are 
input into the calculation, i.e. without introducing further adjustable parameters. 
Comparison with pure nonrelativistic calculations indicates that the relativistic boosts 
play a crucial role even at $Q^2=0$. Moreover, the neutron charge form factor is highly 
sensitive to the details of the GBE model, since it is mostly given by the small 
SU(6)-breaking components of the GBE eigenfunctions. Only for the magnetic response the 
results show an evident deviation from data at $Q^2 > 2$ (GeV/$c)^2$ that leaves room 
for quantitative improvements beyond the PFSA~\cite{GBE3}.

%%%%%%%%%%%%%%%%%%%%%%%%%%%%% NJL model %%%%%%%%%%%%%%%%%%%%%%%%%%%%%%%%%%

\subsection{Baryon Spectrum and Chiral Symmetry Breaking}
\label{subsec:njl}

Chiral symmetry breaking is one of the fundamental QCD features that must be taken into 
account when dealing with the low-energy properties of hadrons. In this section, the 
problem is reconsidered replacing a hamiltonian approach with a phenomenological model 
based on the Nambu Jona-Lasinio (NJL) lagrangian. The strong attractive force between 
quarks in the $J^P=0^+$ channel induces instabilities in the Fock vacuum of massless 
quarks, breaking the chiral symmetry and generating a dynamical quark mass $m^*$ related 
to the vacuum condensate $\langle {\bar q} q \rangle$ and of the same order of magnitude
as the constituent quark mass parameters used in CQM. For increasing density and 
temperature the model reproduces the transition 
from the broken phase to the chiral symmetry restoration. In fact, assuming an 
homogeneous, isotropic quark matter inside the baryon, it is possible to calculate the 
radial dependence of the quark dynamical mass $m^*_a(r)$ for the flavor $a$ in a
Local-Density Approximation (LDA). It reflects the density change when moving away from 
the baryon center: inside the baryon the density (and, consequently, the Fermi momentum) 
is very large, $m^*_a(r)$ is very small and chiral symmetry is restored; getting through 
the baryon surface, the density drops while $m^*_a(r)$ increases until it reaches its 
constituent quark-like asymptotic value, which is determined by the interaction of the
quark with the chirally broken vacuum. 

The energy of the vacuum is ascribed to the filled negative-energy levels of the Dirac 
sea, which put a pressure $p_{\rm vac}$ on the baryon. Viceversa, inside the baryon the
pressure $p_{\rm B,vac}$ is much smaller since the density is very high. The equilibrium 
is given by the counterbalancing pressure $p=p_{\rm vac}-p_{\rm N,vac}$ induced by the 
quarks inside the baryon. This stability condition fixes the size and mass of the 
baryon~\cite{to}. According to different choices of the model parameters, the range 
$\langle r^2\rangle = 0.76 \div 0.80$ fm is obtained for the N radius in very good 
agreement with the experimental datum~\cite{to}. Also the baryon octet spectrum is 
fairly well reproduced within a $\pm 5\%$ deviation, even if the assumed degeneracy 
between $u$ and $d$ flavors prevents from distinguishing the proton from the neutron 
and other isospin multiplets. The deviation from the baryon decuplet spectrum is, 
instead, more evident since no spin-spin force has been taken into account in the
model, while the octet-decuplet splitting is usually attributed to 
the spin-spin interaction. Introducing an effective perturbation, based on the 
spin-spin part of the OGE potential, does not substantially improve the
situation~\cite{to}.

%%%%%%%%%%%%%%%%%%%%%%%%%%%%% DIS %%%%%%%%%%%%%%%%%%%%%%%%%%%%%%%%%%

\section{Deep-Inelastic Scattering}
\label{sec:dis}

Using the optical theorem and the Operator Product Expansion (OPE), the socalled
Cornwall-Norton moments of the structure functions, that enter the hadronic tensor
$W^{\mu\nu}$ for inclusive DIS on the nucleon, can be expanded as  
\begin{equation}
M_n(Q^2) = \int_0^1 dx x^{n-2} F(x,Q^2) = \sum_t \; C_{nt}[\alpha_s(Q^2) {\rm log}(Q^2)] \; 
\; \frac{{\mathcal O}_n^{(t)}}{(Q^2)^{t/2-1}} \; ,
\label{eq:ope}
\end{equation}
where the Wilson coefficients $C_{nt}$ are singular in the limit $Q^2\rightarrow 0$,
while ${\mathcal O}_n^{(t)}$ are matrix elements of regular, symmetric, traceless local
operators. The $C_{nt}$ can be calculated within pQCD at the given order $n$; they
describe the loop corrections that accurately reproduce the observed logarithmic 
violations to scaling in the Bjorken variable $x$. The ${\mathcal O}_n^{(t)}$ contain 
information on the soft physics and are essentially unknown. They are related to 
nonperturbative power corrections through the expansion in twist powers $t$ of $1/Q^2$, 
starting from the leading twist $t=2$. Both $C_{nt}$ and ${\mathcal O}_n^{(t)}$ 
implicitly depend on a renormalization and factorization scales (which will be 
understood in the following), but in any case the OPE gives an automatic and rigorous 
factorization between hard and soft scale physics, even if the proof is limited just to 
inclusive DIS (and $e^+e^-$ annihilation).

The extraction of the socalled higher twists, which in parton language represent 
multiparton correlations, is of delicate importance in determining the full $Q^2$
dependence of the hadronic tensor in the preasymptotic region, since they affect 
accurate extraction from data of the leading-twist parton distributions, which are 
interpreted as parton matrix elements of one-body operators. In the next 
Sec.~\ref{subsec:ope} the issue of higher twist extraction will be discussed for 
the unpolarized and polarized structure functions entering $W^{\mu\nu}$. In 
Sec.~\ref{subsec:pdf} the properties of leading-twist parton distributions will be 
described, with particular emphasys on the socalled transversity distribution, which, 
being a chiral odd object, is unreachable in inclusive DIS and represents the missing 
cornerstone for completing the knowledge of the leading-twist spin distribution of 
quarks inside a hadron. Finally, in Sec.~\ref{subsec:cqm-pdf} I will discuss the 
possibility of linking the parton distributions to the wave functions of constituent 
quarks by properly defining a nonperturbative, low-scale, phenomenological input to 
the Altarelli-Parisi evolution equations.

%%%%%%%%%%%%%%%%%%%%%%%%%%%%% Higher twists %%%%%%%%%%%%%%%%%%%%%%%%%%%%%%%%%%

\subsection{Structure Functions and Higher Twists}
\label{subsec:ope}

In Ref.~\cite{rmIII5}, the moments $M_n(Q^2)$ in Eq.~(\ref{eq:ope}) are obtained by 
fitting the world data for inclusive electron scattering on protons and deuterons in the 
DIS kinematics, including data from SLAC and BCDMS, over the range $1<Q^2<20$ 
(GeV/$c)^2$ and also for invariant masses lower than 2.5 GeV. Since for each $Q^2$ the 
whole range in $x$ is needed, it is necessary to use phenomenological fits that 
interpolate the existing data also in the unexplored kinematics and unavoidably 
introduce a certain amount of uncertainty. Since the OPE holds for inclusive reactions 
only, both elastic and inelastic contribution have been taken into account, the proton 
elastic ones being analytically calculated in terms of its form factors 
$G_E^{\rm p}, G_M^{\rm p}$, while the convolution formula with a realistic N momentum 
distribution is adopted for the deuteron. In Ref.~\cite{rmIII5} it is also remarked that
moments in the Nachtmann variable $\xi$ are preferable, since they allow to filter out 
of $M_n$ spurious higher-twist contributions coming from power-like target mass 
corrections due to the contributions from ${\mathcal O}_{n'}^{(t)}$  operators with 
$n'\neq n$.

The expansion on the right-hand side of Eq.~(\ref{eq:ope}) is approximated up to twist
$t=6$ by using the calculated pQCD result at Next-to-Leading Order (NLO) for the leading 
twist $t=2$, and by introducing effective anomalous dimensions 
$\gamma_{n4}, \gamma_{n6}$ for the twists $t=4,6$, respectively, as input parameters. 
The NLO approximation to the leading twist can be further improved in two ways. A 
Sudakov form factor can be included that accounts for the Soft Gluon Radiation (SGR) at 
any order by using resummation techniques. The correction is required by the Wilson 
coefficients $C_{n2}$ being divergent as log($n$) for $x\rightarrow 1$, where the 
cancellation between singularities produced by virtual and real gluon loops is 
lost~\cite{rmIII5}. Alternatively, at the one-loop level there are power corrections 
deriving from vacuum polarization insertions in gluon lines, called Infra-Red (IR) 
renormalons, that cannot be resummed in Sudakov form factors; at low $Q^2$ they can 
simulate twist corrections, making the extraction of the latter rather ambiguous.

By fitting the Nachtmann moments on the left-hand side of Eq.~(\ref{eq:ope}) with the 
effective anomalous dimension approximation on the right-hand side, it is possible to 
constrain the discussed parameters and extract information on the higher twists. The 
main result is that there is a cancellation between the effective twist-4 and twist-6 
terms, but each one individually is not negligible; this outcome still leaves open the 
question about a possible alternating sign in the twist expansion. It seems, anyway, 
crucial to simultaneously extract all the twists from data. In fact, a comparison with 
the corresponding analysis obtained by fitting the leading twist with a NLO 
parametrization from Gl\"uck, Reya and Vogt (GRV), indicates that small differences at 
leading twist reflect in larger ones at the much smaller higher twists. Most
interestingly, the comparison between the calculated leading-twist $M_n$ for the 
deuteron structure functions and the corresponding GRV parametrization suggests that the 
discrepancy of the latter with data can be eliminated by enhancing the ratio 
$d(x)/u(x)$ for the $u,d$ distributions in the region $x\rightarrow 1$, to
which the higher $n$ moments are mostly sensitive~\cite{rmIII5}. This suggestion is 
consistent with the observed pQCD result ${\textstyle {3\over 7}}$ for the ratio 
$F_2^{\rm n}(x)/F_2^{\rm p}(x)$ in the elastic limit, against an experimental 
(questioned) datum of ${\textstyle {1\over 4}}$. IR renormalons overestimate the
higher-twist contribution with respect to the SGR correction, which is preferable also
because it is analytically calculable.

The above sketched procedure can be extended to inclusive polarized electron 
scattering to extract Nachtmann moments of the polarized structure functions. To this 
purpose, a new phenomenological parametrization of $G_1^{\rm p}(x,Q^2)$ has been 
developed~\cite{rmIII6} by interpolating the world data on proton targets in the DIS 
regime for $x \gtrsim 0.02$ and $1<Q^2 <50$ (GeV/$c)^2$, as well as the available 
results for photo- and electro-production of proton resonances. The parameters are 
constrained to reproduce the experimental value of the Gerasimov-Drell-Hearn (GDH) sum 
rule. Accordingly, the $Q^2$-dependent generalized GDH sum rule is predicted to have a 
zero-crossing point at $Q^2 = 0.16 \pm 0.04$ (GeV/$c)^2$. For the first moment, 
the higher twists are very small. For higher-order moments, particularly sensitive to 
the $x \rightarrow 1$ region, the SGR correction reduces the size of higher twists, 
which, however, remain still relevant for $Q^2$ of few (GeV/$c)^2$, at variance with 
the unpolarized case. This suggests that spin-dependent multi-parton correlations could 
have a bigger impact than the spin-independent ones. Finally, comparing the trend in 
the Nachtmann variable $\xi$ of the interpolated data for $G_1^{\rm p}(\xi,Q^2)$ at 
different low values of $Q^2$ and at $Q^2=20$ (GeV/$c)^2$ in the DIS regime, an 
explicit violation of the Bloom-Gilman duality for polarized electroproduction of 
$\Delta$ on protons is reported.

%%%%%%%%%%%%%%%%%%%%%%%% parton distribution functions %%%%%%%%%%%%%%%%%%%%%%%%

\subsection{Parton Distribution and Fragmentation Functions}
\label{subsec:pdf}

Assuming factorization between hard and soft physics happening at very different scales,
the hadronic tensor $W^{\mu\nu}$ for inclusive DIS on nucleons can be analyzed by the 
approach based on the diagrammatic expansion. The leading-order contribution, both in 
twist and powers of $(\alpha_s {\rm log}(Q^2))$, is given by the well-known handbag 
diagram that involves, in the light-cone gauge $A^+=0$, the soft quark-quark correlation 
function
\begin{equation}
\Phi_a(p) = \int d^4\xi e^{-i\xi \cdot p} \langle P,S| {\overline \psi}_a(\xi)
\psi_a(0) |P,S \rangle \; ,
\label{eq:phi}
\end{equation}
which describes the behaviour of a parton with flavor $a$ and momentum $p$ in a target 
hadron with momentum $P$ and spin $S$ through the hadronic matrix element of a 
leading-twist bilocal operator. Applying the Fierz decomposition to $\Phi_a(p)$, an 
expansion in terms of projections $\Phi_a^{\Gamma}(x)$ onto a complete set of Dirac 
structures $\Gamma$ is obtained, where $x=p^+/P^+$ is the light-cone momentum fraction 
of the parton inside the parent hadron for the dominant light-cone direction "+" induced 
by the electromagnetic interaction (at the considered leading order, the difference 
between $x$ and Bjorken variable will be consistently neglected). The 
$\Gamma=\gamma^+, \gamma^+\gamma_5, i\sigma^{i+}\gamma_5$ projections give the
leading-twist parton model distribution functions: the momentum distribution $f_1(x)$,
the helicity distribution $g_1(x)$ and the transversity distribution $h_1(x)$. Therefore,
while three distributions are needed to reach a complete knowledge of the leading-twist 
momentum and spin structure of the quarks with respect to a preferred longitudinal 
direction, only two of them, $f_1$ and $g_1$, have been extensively studied through 
measurements of the corresponding unpolarized and longitudinally polarized structure 
functions, respectively. The reason is that the transversity $h_1$ is related to 
helicity flipping mechanisms; hence, it is a chiral-odd object and requires the 
occurrence of another chiral-odd partner in any observable to be measured. In inclusive 
DIS at leading twist, there are no mechanisms providing such partners because chirality 
is preserved in hard pQCD. Nevertheless, $h_1$ is a very interesting object. I forward 
the interested reader to the very comprehensive report of Ref.~\cite{h1rep} on the 
$h_1$ properties. Here, it is sufficient to mention that its integral measures the 
tensor charge, whose determination would give access to chiral-odd operators in QCD and, 
consequently, to the role of chiral symmetry in the hadron structure. Because of the 
mismatch in helicities, $h_1$ has a non-singlet evolution disconnected from gluons, hence
more smooth than the one of $g_1$. In the hadron rest frame, $h_1$ and $g_1$ would be 
perfectly equivalent because of the rotational symmetry of the system. However, the 
distributions are defined after the system is boosted along a preferred longitudinal 
direction induced by the electromagnetic interaction, thus breaking the rotational 
invariance. Therefore, the difference between $h_1$ and $g_1$ tells us also about the 
relativistic nature of the dynamics of quarks inside the hadron. 

Originally, extraction of $h_1$ was proposed by measuring spin asymmetries in 
Drell-Yan reactions with two colliding transversely polarized protons; however, even 
if the process occurs at leading twist, the probability of finding transversely 
polarized antiquarks is presumable much smaller than for quarks. In DIS, semi-inclusive 
processes need to be considered to provide the necessary chiral-odd partner through the 
quark hadronization into one or several detected hadrons. Similarly to 
Eq.~(\ref{eq:phi}), the corresponding soft correlation function can be parametrized in 
terms of the socalled fragmentation functions, which also have at leading twist a 
probabilistic interpretation and are interesting quantities by themselves since they 
bear witness of how confinement arises. However, the twist analysis reveals that for 
detected spinless hadrons chiral-odd fragmentation functions appear only at the 
twist-three level. The price to pay for having such objects at leading twist is to 
detect baryons, such as the $\Lambda$ particle, in socalled double-spin asymmetries 
where both the target and the $\Lambda$ are polarized. Before commenting on this option, 
I will discuss the simpler alternative of the single-spin asymmetry, where only the 
target is polarized and final mesons (basically pions) are abundantly measured in the 
detector. The key observation is that if memory of the transverse quark dynamics is kept 
inside the various distributions, it is possible to build an azimuthal asymmetry
related to the socalled Collins angle $\sin \phi_C \propto {\bf P}_h \times {\bf k} 
\cdot {\bf S}_T$, where $P_h$ is the detected hadron momentum with a nonvanishing 
transverse component with respect to the jet axis and ${\bf k}$ is the fragmenting quark
momentum with transverse polarization ${\bf S}_T$~\cite{ca1}. By properly weighting the
single-spin asymmetry obtained by flipping the transverse polarization of a nucleon
target, it is possible to isolate $h_1$ at leading twist by means of the first moment
of the socalled Collins function $H_1^\perp (z,{\bf k}_T)$, where $z=P_h^-/k^-$ is the 
light-cone momentum fraction of the fragmenting quark carried by the detected hadron. 
This chiral-odd fragmentation function is given by the probability difference of quarks 
with opposite transverse polarization to fragment into an unpolarized hadron; it can be 
interpreted as a sort of analyzing power of the quark polarization into the orbital 
angular motion of the detected hadron. 

The Collins effect has been measured at Fermilab using the reaction $pp^\uparrow
\rightarrow \pi X$ and by the HERMES collaboration using the DIS $e\vec p 
\rightarrow e'\pi X$, where the longitudinal polarization of the target along the beam
direction has a small transverse component with respect to the direction of the virtual
photon~\cite{ca2}. Because of such an effective small polarization and of an asymmetry
dilution due to NLO corrections introduced by SGR techniques, it came recently as a 
surprise that the measured asymmetry for all detected isospin states of the pion has a 
remarkable absolute size of as much as 5\%. At present, there is an ongoing discussion 
about the issue if factorization is under control or if other possible mechanisms could 
generate a nonvanishing single-spin asymmetry that could make the $h_1$ extraction 
questionable. I refer the interested reader to Ref.~\cite{ca-talk} and references
therein. Here, I will just mention the work of Ref.~\cite{ca3}, where the 
Weinberg's idea of a generalized time-reversal operator is applied to a flavor 
multiplet of quarks interacting with chiral fields. By mixing up the flavors inside 
the multiplet on top of the usual time-reversal transformation, the internal N chiral 
dynamics generates ${\bf k}_T$-dependent distributions functions that would be 
forbidden by the standard time-reversal operator, but are perfectly legitimate under 
the new nonstandard one. This would open the possibility of a larger class of 
transverse single-spin asymmetries. Because of the supposed universality of such naive 
$T$-odd distributions, the same possibility should be open also in semi-inclusive 
lepton DIS, thus giving a possible interpretation of the HERMES data. 

A very promising alternative to the Collins effect is represented by the single-spin
asymmetry generated by the correlation between the transverse polarization of the
fragmenting quark and the orientation of a $\pi^+\pi^-$ pair inside the jet in
semi-inclusive production of two leading pions. The effect is proportional to the
azimuthal angle $\sin \phi \propto {\bf P}_{\pi^+} \times {\bf P}_{\pi^-} \cdot 
{\bf S}_T$ and can be interpreted as due to the interference between different relative
partial waves of the pion pair with different phases. By generalizing the soft 
correlation function to this case, a leading-twist analysis reveals that a new
class of chiral-odd (and naive $T$-odd) fragmentation functions arises, the socalled 
Interference Fragmentation Functions (IFF), that depend in a complicated way upon the
kinematics and the ``geometry'' of the reaction~\cite{IFF1}. However, in DIS a
single-spin asymmetry can isolate $h_1$ at leading twist through 
$H_1^{<\kern -0.3 em{\scriptscriptstyle )}}$, which is an IFF sensitive to the 
transverse component of the pair relative momentum~\cite{IFF2}. The asymmetry is based 
on a collinear factorization theorem, where the ${\bf k}_T$ dependence can be integrated 
away leading to a cancellation of soft gluon radiation effect, at variance with the
Collins function; an analysis beyond tree level seems, therefore, straightforward. The 
$H_1^{<\kern -0.3 em{\scriptscriptstyle )}}$ could be extracted, in principle, from the
corresponding $e^+e^-$ annihilation. At present, its knowledge relies upon models. In 
the context of the spectator model approximation, the single-spin asymmetry for the DIS 
$ep^\uparrow \rightarrow e'(\pi^+\pi^-)X$ has been computed by modelling the IFF as 
originating from the interference between an $s$-wave direct $\pi\pi$ production and a 
$p$-wave decay of the $\rho$ resonance, leading to an asymmetry as large as 
1.5\%~\cite{IFF2}. 

I will now go back to the case of semi-inclusive production of the $\Lambda$ particle. 
For the semi-inclusive production of transversely polarized $\Lambda^\uparrow$, a 
double-spin asymmetry depends at leading twist upon the product $h_1 H_1$, where $H_1$ 
is a chiral-odd (naive $T$-even) fragmentation function related to the probability of a 
transversely polarized quark to fragment into a transversely polarized baryon. A 
sizeable asymmetry has been observed at Fermilab, but the low values of 
$\Lambda^\uparrow$ transverse momentum explored prevent from using any
factorization in the analysis of the cross section, thus posing questions on the $h_1$
extraction from these data. Moreover, modelling $H_1$ is a challenge since
the mechanism for the polarization transfer from the quark to the $\Lambda$ is still
unknown. The data collected at Fermilab for the unpolarized $\Lambda^\uparrow$ 
production ($pp\rightarrow \Lambda^\uparrow X$) also show a very
large asymmetry around 20\%, while pQCD predicts an almost vanishing effect if the
process is assumed collinear. Since the $\Lambda$ polarization seems roughly
independent from the target, a possible interpretation is given by a new
nonperturbative spin effect that produces an asymmetry proportional to the
azimuthal angle $\sin \phi \propto {\bf k} \times {\bf k}_T \cdot {\bf S}_\Lambda$. It 
is parametrized by the polarized Fragmentation Function (polFF) $D_{1T}^\perp$, which
is related to the probability of an unpolarized quark to fragment into a transversely
polarized hadron~\cite{ca4}. The $D_{1T}^\perp$ is naive $T$-odd but chiral even; 
therefore, it can be easily detected also in $\Lambda$ productions with charged-current 
processes. Assuming a factorization theorem and charge conjugation invariance to hold, 
a fit to the data for the $p{\rm Be}\rightarrow \Lambda^\uparrow X$ and 
$p{\rm Be}\rightarrow {\overline \Lambda}^\uparrow X$ can be achieved with a simple
parametrization for $D_{1T}^\perp$. The known trends of a large negative $\Lambda$
polarization, increasing with $x_F=p_\Lambda/p_{\rm beam}$, of its puzzling flat
dependence on $\Lambda$ transverse momenta beyond 1 GeV/$c$, and of an almost
vanishing $\overline \Lambda$ polarization, are reasonably reproduced~\cite{ca4}. 
However, the Fermilab data again prevent any interpretation in terms of factorization. 
Therefore, the extracted $D_{1T}^\perp$ cannot be used in the 
$ep\rightarrow e'\Lambda^\uparrow X$ reaction, where only speculations based on 
different scenarios and different choices of parameters are possible, at 
present~\cite{ca5}. 

When the $\Lambda$ and $\overline \Lambda$ are longitudinally polarized, new
possibilities are available for deepening the knowledge of parton
distributions~\cite{ca6}. Working at LO in pQCD, the cross section for the process 
$\vec l \vec {\rm N} \rightarrow l' \vec \Lambda X$ has eight helicity components 
which can be reduced to just four by $P$-invariance and helicity conservation. Assuming 
charge-conjugation and isospin invariance, linear combinations of these four components 
for p and n targets allow to extract information on the polarized strange quark density 
$g_1^s(x)+g_1^{\overline s}(x)$ and on the validity of the hypothesis 
$f_1^s(x)=f_1^{\overline s}(x)$ and $g_1^s(x)=g_1^{\overline s}(x)$, that are not
reachable in unpolarized DIS~\cite{ca7}. By generalizing the analysis also to 
charge-current processes and $e^+e^-$ annihilations leading to $\vec \Lambda$ and 
$\vec {\overline \Lambda}$, various parametrizations can be considered for the 
unpolarized $D_1^a$ and longitudinally polarized $G_1^a$ fragmentation functions for 
quark flavor $a$ to fit recent data from various laboratories~\cite{ca8}, in order to 
test various scenarios according to different assumptions in the flavor SU(3) 
decomposition.

%%%%%%%%%%%%%%%%%%%%%%%% cqm-pdf evolution %%%%%%%%%%%%%%%%%%%%%%%%

\subsection{From Constituent Quarks to Parton Distributions}
\label{subsec:cqm-pdf}

It has already been remarked in Sec.~\ref{subsec:ope} that the ratio 
$F_2^{\rm n}(x)/F_2^{\rm p}(x)$ in the limit $x\rightarrow 1$ is an observable very
sensitive to the symmetry properties of the underlying theory, which can influence the
parton distributions, in particular the ratio $d(x)/u(x)$. When studying the quark
distributions, usually, the second moment of the valence quark distribution is 
evolved back from a low-energy parametrization to a low-energy scale $\mu_0$ where the 
short-range perturbative contribution (related to the gluon and sea quarks) is 
negligible and the valence distribution becomes dominant. Then, LO or NLO evolution is 
applied up to the selected experimental scale and the structure functions are deduced in 
the partonic description of quark distributions. Since the parton model emerges in the 
Bjorken limit because of light-cone dominance, it can be developed in the hadron rest 
frame using the light-cone formalism. The low-energy parametrization of the polarized 
valence quark distribution $q_\lambda^{\rm val}(x,\mu_0^2)$ can then be replaced by the 
light-cone momentum density $n_\lambda$ of the parton with helicity $\lambda$. The 
latter can be calculated in the light-front formulation of a relativistic interacting 
three-body sistem (see Sec.~\ref{subsec:lf}). Since, moreover, the usual scale $\mu_0$ 
turns out to be very close to the constituent quark mass, the partons in the hadron rest 
frame can be identified with the constituent quarks. Thus, the low-energy input to the 
evolution equations can be related to the CQM N wave functions, that are deduced by 
satisfying covariance and the Pauli principle. In this way, it is possible to test at the
high DIS scale the mechanisms proposed for SU(6) breaking at low scale. The issue has 
been debated since long time, but only recently in Ref.~\cite{pv1} a correct treatment 
of the effects related to covariance and Pauli principle have been accounted for. For 
the range $0.3 \lesssim x \lesssim 0.7$, where the valence quark picture dominates and 
the approach is meaningful, the ratio $F_2^{\rm n}(x)/F_2^{\rm p}(x)$ has been 
investigated using the two potentials $V_{\rm OGE}$ and $V_{\rm GBE}$. Both interactions 
give asymptotic values different from the SU(6) result ${\textstyle {2\over 3}}$, but 
fail to reproduce the high-$x$ trend of the CTEQ5 parametrization fitting the data, even 
if the ratio shows a rather weak sensitivity to QCD radiative effects introduced by 
evolution. The disagreement emerges only if the effects due to Pauli principle are 
properly included~\cite{pv1}.  

The same technique can be applied to the study of the orbital angular momentum of the
hadron constituents~\cite{tn1}. The interest in this observable is dictated by the 
common belief that the N total angular momentum of ${\textstyle {1\over 2}}$ must be 
shared among its constituents as 
${\textstyle {1\over 2}}={\textstyle {1\over 2}}\Delta\Sigma+\Delta g+L_q+L_g$, where
$\Delta\Sigma, \Delta g$ are the helicities of quarks and gluons, respectively, and
$L_q, L_g$ are the quark and gluon orbital angular momenta. A gauge invariant definition 
of $L_q$ and $L_g$ involves the socalled Generalized Parton Distributions (GPD), that 
will be addressed in the next Sec.~\ref{sec:gpd}. Here, a low-energy model calculation 
of $L_q$ and $\Delta\Sigma$ is obtained by calculating matrix elements of the 
$\sum_i -i({\bf k}_i\times{\bf \nabla}_i)$ and $(n_\lambda-n_{-\lambda})$ operators on 
SU(6) wave functions obtained in LF description of the hypercentral CQM described in 
Sec.~\ref{sec:cqm}~\cite{tn1}. Melosh rotations of the quark spins generate a 
nonvanishing $L_q$ at the input scale $\mu_0$ even if the N state is in a $S$ wave. The 
model verifies the N angular momentum sum rule, which reads 
${\textstyle {1\over 2}}\Delta\Sigma (\mu_0^2) + L_q(\mu_0^2) = 0.228 + 0.272 = 0.5$. 
Then, 
evolution mixes up the two contributions, but the effect of Melosh rotations and of high 
momentum components, generated by $V_{\rm hyp}$, survives at large $x$. Evolution 
produces also $L_g(Q^2)\neq 0$, even if $L_g(\mu_0^2)=0$. The values obtained in the 
range $1<Q^2<20$ (GeV/$c)^2$ are compatible with the ones obtained by QCD sum rules and 
lattice calculations~\cite{tn1}.

%%%%%%%%%%%%%%%%%%%%%%%%%%%%% GPD %%%%%%%%%%%%%%%%%%%%%%%%%%%%%%%%%%

\section{Generalized Parton Distributions}
\label{sec:gpd}

As already anticipated in the previous subsection, Generalized Parton Distributions
(GPD) represent a possible tool to address the orbital angular momentum of quarks
inside hadrons in a gauge-invariant way. Intuitively, GPD are recovered by considering 
the handbag diagram of inclusive DIS and pushing it to the off-diagonal components in 
the hadron and parton legs, as it may happen for the leading-order diagram of Deeply
Virtual Compton Scattering (DVCS) or of hard exclusive Meson Production (DVMP). For very 
large $Q^2$ of the entering virtual photon, for large cm energy $s$ and small momentum
transfer $t$, a factorization theorem can be rigorously proven between the hard 
electromagnetic interaction and the soft physics parametrized in terms of a quark-quark 
correlation function, that generalized the one in Eq.~(\ref{eq:phi}). The latter can be
expanded in the relevant Dirac structure displaying, at leading twist, two unpolarized
and two polarized helicity-conserving and helicity-flipping GPD. Because of the 
factorization theorem, the GPD formalism represents a unifying framework for whole 
classes of inclusive and exclusive processes. In fact, in the forward limit the usual 
parton distributions are recovered. The first moments in $x$ produce the 
electromagnetic form factors $F_1(t), F_2(t)$; if the final hadrons are N resonances, 
the transition form factors are also obtained. From GPD spatial transverse distributions
can be defined that give information on the localization of partons inside hadrons.
Finally, at $t=0$ the $m$-th moments in $x$ are even polynomials of degree $m$ in the 
skewedness parameter $\xi$, which is related to the mismatch in the longitudinal hadron
momentum. This constraint, called polynomiality property, is a very general one, since 
it follows from Lorentz invariance. When applied to the unpolarized GPD for $m=2$, a 
gauge invariant definition of the quark total angular momentum is obtained. 

The experimental extraction of GPD is a very challenging task. Recently, data have been
obtained for the beam spin and charge asymmetry, and for DVCS, at the DESY and TJNAF 
laboratories; these measurements, however, allow for mapping GPD only in a very limited
part of the phase space. Therefore, at present their understanding relies mostly on 
models. In the literature, there are basically two different approaches. The first one 
is a phenomenological construction based on a reduction formula where the GPD are 
related to the parton distributions by factorizing the $t$ dependence due to the 
electromagnetic form factors. The latter approach is based on a direct calculation with 
specific dynamical assumptions. In Ref.~\cite{pv2}, the chiral quark-soliton  model is 
considered, based on the instanton realization of the nonperturbative QCD vacuum, and a 
further proof of its internal consistency is given by analytically checking that the 
polynomiality property is satisfied. As a byproduct, explicit expressions are obtained 
for all the coefficients of the $m$-th order polynomial; in particular, the 
highest-order coefficient gives the $m$-th moment of the socalled $D$-term, which gives 
the behaviour of the GPD in the nonvalence domain $-\xi < x < \xi$. 

By expanding the N state upon valence light-cone wave functions in the Fock space for 
three partons, the off-diagonal soft correlation function can be represented as an 
overlap of the valence 
light-cone wave functions themselves. Very similarly to Sec.~\ref{subsec:cqm-pdf}, the 
GPD can then be represented in terms of the off-diagonal light-cone density matrix, 
which, in turn, can be calculated by solving the eigenvalue problem for some 
quark-quark potential in a CQM and then by transforming the solutions to the 
light-cone via the LF boosts~\cite{pv3}. The correct covariance of the approach grants 
the good properties of the GPD, namely recovering the parton distributions with the 
correct support in the forward limit and automatically fulfilling the particle number 
and momentum sum rules. However, since the GPD are calculated from valence light-cone 
wave functions, they are defined only in the region $\xi \leq x \leq 1$; 
therefore, their first moment gives the N electromagnetic form factors only in the 
limit $\xi\rightarrow 0$. In Ref.~\cite{pv3}, the leading-twist unpolarized GPD are 
calculated starting from the $V_{\rm hyp}$ and $V_{\rm GBE}$ potentials described in 
Sec.~\ref{sec:cqm}. No particular difference is obtained with the two choices, despite
the fact the former is SU(6) symmetric and the latter contains SU(6)-breaking terms.
Remarkably, the unpolarized helicity-flipping GPD is completely determined by the effect 
of Melosh rotations in boosting the CQM wave functions on the light-cone. A strong $t$ 
and a weak $\xi$ dependence are observed.

\end{document}